\documentclass[aps,notitlepage,showpacs]{revtex4}
\usepackage{graphicx}
\usepackage{float}
\usepackage[T1]{fontenc}
\usepackage[latin1]{inputenc}
\usepackage{amssymb}
\include{epsf}
\epsfverbosetrue

\newcommand{\ket}[1]{\ensuremath{|#1\rangle}}			
\newcommand{\iprod}[2]{\ensuremath{\langle#1|#2\rangle}}	
\newcommand{\bea}{\begin{eqnarray}}
\newcommand{\eea}{\end{eqnarray}}
\newcommand{\ts}{\vspace*{.2cm}}	

\begin{document}
\large

\title{Geometrical entanglement of highly symmetric multipartite states and the Schmidt decomposition}

\author{D.~Buhr$^*$, M.E.~Carrington$^\dagger$, T.~Fugleberg$^\dagger$, R. Kobes$^*$, G. Kunstatter$^*$, D. McGillis$^\dagger$, C. Pugh$^\dagger$ and D. Ryckman$^*$}

\affiliation{$^\dagger$ Brandon University, Brandon, Manitoba, R7A 6A9 Canada, 
and Winnipeg Institute for Theoretical Physics,
Winnipeg, Manitoba, Canada}

\affiliation{$^*$ Physics Department, University of Winnipeg, Winnipeg, MB, R3B 2E9, Canada and Winnipeg Institute for Theoretical Physics, Winnipeg, Manitoba, Canada}

\begin{abstract}
In a previous paper  we examined a geometric measure of entanglement based on the minimum distance between the entangled target state of interest and the space of unnormalized product states. 
Here we present a detailed study of this entanglement measure for $n-$qubit target states that are invariant under the  permutation of any two qubits. We analytically obtain the permutation invariant unnormalized direct product states that extremize the distance function. We then solve for the Hessian to show that, up to the action of trivial symmetries, the solutions correspond to local minima of the distance function. In addition, we show that the conditions that determine the extremal solutions for general target states can be obtained directly by parametrizing the product states via their Schmidt decomposition. 
\end{abstract}
\date{\today}

\maketitle

\section{Introduction}
\label{sectionIntro}

Given the vital role that quantum entanglement
\cite{review1,review2,review3} is thought to play in quantum computing \cite{nielsen} and its role in quantum phase transitions\cite{amico}, it is important to have at one's disposal a quantitative measure of entanglement. General conditions that must be satisfied by useful entanglement measures were presented in \cite{quantify}. In that paper a class of good measures was constructed based on the notion of minimum distance between the state in question (i.e. the target state) and the nearest disentangled state. There has been much work on so-called  geometric measures of entanglement based on this distance to the nearest product
state \cite{g1,g2,ex1,witte,ozawa}. { Geometrical entanglement was applied to multipartite (3 qubit) systems in \cite{g3} and related to the notion of entanglement witness in \cite{bertlmann,witness}. There has more recently been a resurgence of interest in various aspects of geometrical entanglement, including formalism, generalizations and calculations for specific systems (see \cite{Leinass,g4,hierarchies,accessible,orus,three qubit,orus08,orus10,estimation,hubener}, for a partial list). } Explicit calculations of such measures for multipartite systems are generally very complicated and difficult to compute, except for target states with a high degree of symmetry.

In a previous paper \cite{passante}, we studied a variation of the geometric measure of entanglement based on the minimum distance between an 
{\bf unnormalised} product state and a target entangled state\footnote{To the best of our knowledge, a geometrical measure using unnormalized product states was first considered by Schulman and Mozyrsky \cite{schulman}. { Leinass {\it et al} considered a relaxation of the normalization condition in \cite{Leinass}.}}. The equations determining the (unnormalized) geometrical entanglement of a multipartite state were presented for a general target state. { As shown in detail in \cite{passante} and reviewed at the end of the next section, there is a simple geometrical relationship between the geometrical entanglement calculated  using unnormalized product states and those that are normalized.} Although the equations are non-linear and hence difficult to solve in general, for target states with a large degree of symmetry it is possible to find analytic solutions that extremize the distance function. 

For a given solution to the variational equations to yield a meaningful measure of entanglement, it  must be a local minimum of the distance function\footnote{ Ideally it should be a global minimum, but this is more difficult to  determine and beyond the scope of the present paper.}, { and not a local maximum, or an inflection point in any direction within the parameter space. This information is provided by the Hessian for the system, which is the matrix of second variational derivatives evaluated at the extremum. In particular, all nonzero eigenvalues of the Hessian must be positive.} Moreover, any zero eigenvalues should correspond to trivial symmetries of the system. 

In this paper, we address this issue in the context of a maximally symmetric, permutation invariant target state consisting of $q$ qubits \footnote{  The geometrical entanglement of states in Lipkin-Meshkov-Glick model with large (permutation) symmetry was calculated in \cite{orus10}. The first calculation of entanglement of the ground state in this model was done in \cite{LMG}.}. In addition, we are able able to show quite generally that the geometric measure of entanglement has a direct interpretation in terms of a Schmidt decomposition of the multi-partite product states that enter the distance function. This result extends the connection with the Schmidt decomposition that was established in \cite{passante} for bi-partite systems. 

The paper is organized as follows: in the next Section we establish notation and review the geometric entanglement measure and resultant extrema conditions introduced in \cite{passante}. Section \ref{hessianSection} presents a general discussion of the Hessian and evaluates it  for a maximally symmetric target state of $q$ qubits. Section \ref{esystemSection} presents the eigenvalues and eigenvectors of the Hessian (the explicit derivation is delegated to Appendix A) while Section \ref{interSection} discusses the physical/geometrical interpretation of the eigenvectors in terms of their action on the space of product states.  In Section \ref{AnaSection} we use the results of the previous sections to derive the analytic solution for the minimum distance and verify explicitly that it is a local minimum. In Appendix B, we illustrate that while our symmetric ansatz for the product states does solve the extremization conditions for a target state with less symmetry, it does not in that case correspond to a local minimum. Section \ref{schmidtSection} describes the our distance measure in terms of a parametrization using the Schmidt decomposition of the product states. We end with Conclusions and prospects for future work.

\section{Notation and Summary of Previous Results}

We start by defining our notation. We consider a system of $q$ qubits. The dimension of the corresponding Hilbert space is $n=2^q$. We decompose the system into a set of $q$ subsystems, each of dimension 2. The subsystems are labelled $A,B,C,\cdots$ An arbitrary set of basis states of system $A$ is labelled $|i\rangle$, the basis states of $B$ are $|j\rangle$, the basis states of $C$ are $|k\rangle$, etc, and $\{i,j,k,\cdots\}\in\{0,1\}$. Using this notation we write:
\bea
\ket{A}=a_i\ket{i}\,,~~\ket{B}=b_j\ket{j}\,,~~\ket{C}=c_k\ket{k}\,,~~\dots
\eea
where the summation convention is implied. The coefficients $a_i,b_j,..$ are in general complex but we will henceforth consider them to be real for simplicity. Much of the following can be generalized to complex coefficients in a straightforward manner.

The wave-function of an arbitrary normalised entangled pure state is written:
\bea
\label{psi}
\ket{\psi} = \chi_{ijk\cdots} \ket{i}\otimes \ket{j}\otimes \ket{k}\cdots~,~~~\iprod{\psi}{\psi}=1\,.
\eea
We consider a general product state of the form:
\begin{equation}
\label{phi}
 \ket{\phi} = \ket{A}\otimes\ket{B} \otimes \ket{C}\otimes\ldots
   = a_i\ket{i} \otimes 
   b_j\ket{j} \otimes
   c_k\ket{k} \otimes\ldots = \phi_{ijk\cdots}\ket{i}\otimes \ket{j}\otimes \ket{k}\cdots
\end{equation}
Note that the state $\ket{\phi}$ is not assumed to
be normalised:
\bea
\label{norm}
\iprod{\phi}{\phi} = N_AN_BN_C\ldots\,, ~~
N_A =\iprod{A}{A}=a_i^*a_i\,,~~
N_B =\iprod{B}{B} = b_j^*b_j\,,~~\cdots
\eea
The distance between the states $\ket{\psi}$ and $\ket{\phi}$ is \cite{passante}:
\bea
\label{DsqDefn}
D^{2}=\langle \psi-\phi|\psi-\phi \rangle = \left|a_{i}b_{j}c_{k}\ldots-\chi_{ijk\ldots}\right|                     ^{2}\,.
\eea
This distance function depends on the $2q$ parameters $\{a_i,b_j,c_k\cdots\}$.
Taking the first derivatives with respect to the coefficients $a_{i}$ and setting the result to zero produces 2 equations which can be written:
\bea
\label{der1}
\frac{\partial}{\partial a_{i}}D^{2} & = & 2\left(a_{i}b_{j}c_{k}\ldots-\chi_{ijk\ldots}^{}\right)b_{j}c_{k}\ldots
=  2\left(a_{i}\, b_{j}^{2}\,c_{k}^{2}\right)\ldots -2\left(\chi_{ijk\ldots}^{}b_{j}c_{k}\ldots\right) = 0\,,~~i\in\{0,1\}\,.
 \eea
In the same way, we can take derivatives with respect to the variables $b_j$, $c_k$, $\cdots$ and set the resulting expressions to zero. We obtain a set of $2q$ equations which depend on the $2q$ variables $\{a_i,b_j,c_k\cdots\}$. Substituting these solutions back into the distance function gives the condition that must be satisfied by an extremal solution \cite{passante}:
\bea
D_c^2 = 1-N_AN_BN_C\ldots \,.
\label{cdistance}
\eea
We can write this condition in terms of a critical angle $\theta_C$ which is defined as the angle between $\ket{\psi}$ and $\ket{\phi}$ at the extrema: 
\bea
\label{cangle}
&& D_c^2= 1-\cos^2\theta_c\,,\\
&& \cos\theta_c = \left.
\frac{ \iprod{\psi}{\phi}}{\sqrt{\iprod{\phi}{\phi}}\sqrt{\iprod{\psi}{\psi}}}
\right|_{\rm critical} =
\sqrt{N_AN_BN_C\ldots}\,.\nonumber
\eea
{ The relationship of the unnormalized measure to the measure using a normalized product state is revealed by noting as in \cite{passante} that the minimization of the distance used to obtain the latter can be expressed in terms of the same variational principle as the former, but with the normalization condition on the product states imposed using a lagrange multiplier. The corresponding result for the distance to a closest normalized product state is\cite{passante}:
 \bea
 D_N^2=\langle\phi_N-\psi|\phi_N-\psi\rangle=D_c^2 + (1-\sqrt{\langle\phi|\phi\rangle})^2=2D_C^2.
 \eea
The final relationship above follows from a simple geometrical identity. The expectation therefore is that the two measures would in most cases be physically equivalent. One advantage of the unnormalized measure is that the variational equations without the normalization contraint are in principle somewhat simpler (although still non-linear and difficult to solve). }

\section{The Hessian}
\label{hessianSection}


The Hessian is the $2q\times 2q$ matrix of second derivatives of the distance function with respect to the parameters $\{a_i,b_j,c_k\cdots\}$, evaluated at the given extremum. { Denote by $x^a$ the complete set of numbers $\{a^i,b^j,...\}$ parametrizing the unnormalized product states. Condition (\ref{der1}) can then be written:
\bea
\left.\frac{\partial D^2}{\partial x^a}\right|_{\overline{x}}=0
\label{extremal}
\eea
for some soluton $\overline{x}$. The value of the distance function at the extremum is $D^2(\overline{x})$. If one moves away from the minimum by a small displacement $\delta x^a$, the value of the distance function at the minimum changes by:
\bea
\delta D^2 = \left.\frac{\partial^2 D^2}{\partial x^a \partial x^b}\right|_{\overline{x}}\delta x^a\delta x^b+ O(\delta x)^3 
\eea 
where we have used the extremal condition (\ref{extremal}) to eliminate the term linear in the variation. The Hessian:
\bea
H_{ab}\equiv \left.\frac{\partial^2 D^2}{\partial x^a \partial x^b}\right|_{\overline{x}}
\eea
provides information about how the distance functions changes as one moves away from the extremum in parameter space. Specifically, the eigenvectors of the Hessian indicate the directions in parameter space in the neighbourhood of the extremum in which the distance function is increasing (positive eigenvalue), decreasing (negative eigenvalue) or unchanging (zero eigenvalue). For the extremum to be a local minimum, all eigenvalues must be positive, apart from the zero eigenvalues associated with symmetries of the system. We now evaluate the general Hessian for the system under consideration. }

Diagonal terms
of the Hessian have the form (no summation on $i$):
\bea
\label{ON}
\frac{\partial^{2}}{\partial a_{i}^{2}}D^{2}=2\left(b_{j}^{2}c_{k}^{2}\cdots \right)=:\tau_{a_i}\,.
\eea
The following off diagonal terms vanish identically:
\bea
\frac{\partial^{2}}{\partial a_{0}\partial a_{1}}D^{2}=\frac{\partial^{2}}{\partial a_{1}\partial a_{0}}D^{2}=\frac{\partial^{2}}{\partial b_{0}\partial b_{1}}D^{2}=\cdots=0\,.
\eea
All remaining terms have the form:
\bea
\label{OFF}
\frac{\partial^{2}}{\partial a_{i}\partial b_{j}}D^{2}=4a_{i}b_{j}c_{k}^{2} d_l^2\cdots-2\left(\chi_{ijk\ldots}^{}c_{k}d_{l}\cdots\right)=:\gamma_{a_ib_j}\,.
\eea
Using this notation the Hessian can be written:
\bea
\label{hessGeneral}
H=\left(\begin{array}{ccccccc}
\tau_{a_0} & 0 & \gamma_{a_{0}b_{0}} & \gamma_{a_{0}b_{1}} & \gamma_{a_{0}c_{0}} & \gamma_{a_{0}c_{1}} & \cdots\\
0 & \tau_{a_1} & \gamma_{a_{1}b_{0}} & \gamma_{a_{1}b_{1}} & \gamma_{a_{1}c_{0}} & \gamma_{a_{1}c_{1}} & \cdots\\
\gamma_{b_{0}a_{0}} & \gamma_{b_{0}a_{1}} & \tau_{b_0} & 0 & \gamma_{b_{0}c_{0}} & \gamma_{b_{0}c_{1}} & \cdots\\
\gamma_{b_{1}a_{0}} & \gamma_{b_{1}a_{1}} & 0 & \tau_{b_1} & \gamma_{b_{1}c_{0}} & \gamma_{b_{1}c_{1}} & \cdots\\
\gamma_{c_{0}a_{0}} & \gamma_{c_{0}a_{1}} & \gamma_{c_{0}b_{0}} & \gamma_{c_{0}b_{1}} & \tau_{c_0} & 0 & \cdots\\
\gamma_{c_{1}a_{0}} & \gamma_{c_{1}a_{1}} & \gamma_{c_{1}b_{0}} & \gamma_{c_{1}b_{1}} & 0 & \tau_{c_1} & \cdots\\
\vdots & \vdots & \vdots & \vdots & \vdots & \vdots & \ddots\end{array}\right)\,.
\eea

We wish to study the unnormalized geometrical entanglement of target states that are permutation invariant. We make the {\it ansatz} that the  closest product state has the same symmetry and restrict consideration to evaluating the Hessian for product states that , like the target state, are maximally permutational invariant: 
\begin{eqnarray}
\label{randyRule}
a_0 &=& b_0 = \cdots = \alpha_0\,, \nonumber\\
a_1 &=& b_1 = \cdots = \alpha_1\,,
\end{eqnarray}
which gives:
\bea
\label{ansatzCond}
N_A=N_B=N_c=\dots = N:=\alpha_0^2+\alpha_1^2\,.
\eea

{ It is important to note that in general the closest product state may not necessarily have the same number of symmetries as the target state. It was proven in \cite{hubener} that the closest {\bf normalized} product state to a permutationally invariant target state is also permutationally invariant\footnote{This is not true, for example, for states that are translationally invariant \cite{hubener,orus10}.}. This has not been proven for unnormalized product states, so that (\ref{randyRule}) is for the moment merely an {\it ansatz}. In order for it to provide a correct measure of the geometrical entanglement, the {\it ansatz} must not only provide an extremal solution to the variational equations, but as discussed above it must correspond to a local minimum. In section \ref{AnaSection} we show below that for the permutation invariant target states under consideration, the eigenvalues of the Hessian evaluated at extrema of the form (\ref{randyRule}) are indeed positive. This will prove  that maximally symmetric unormalized product states do provide a local minimum to the distance function. It is beyond the scope of the present paper to prove that these are in fact global minima.}

Equation (\ref{der1}) and the corresponding equations obtained by differentiating respect to the variables $b_j$, $c_k$, $\cdots$ form a set of $2q$ equations which depend on the $2q$ variables $\{a_i,b_j,c_k\cdots\}$. When we use the ansatz in Eq. (\ref{randyRule}), $2(q-1)$ of these equations are automatically satisfied, and the remaining two equations determine the values of the two paramaters $\alpha_0$ and $\alpha_1$ which correspond to an extremal solution. 

Using (\ref{randyRule}), the diagonal terms in (\ref{ON}) are all equal:
\bea
\label{on}
\tau_{x_i}=2N^{q-1} =:\tau\,,~~x\in\{a,b,c,\cdots\}\,.
\eea
Each of the off-diagonal terms in  (\ref{OFF}) is equal to one of three terms which we denote $\gamma_{00}$, $\gamma_{01}$ or $\gamma_{11}$ (because of the symmetry of target state (Eq. (\ref{randyRule})), we have $\gamma_{10}=\gamma_{01}$).
We can write all three of these terms collectively as (no summation on $i$ and $j$):
\bea
\label{off}
\gamma_{ij} = 4\alpha_{0}^{2-(i+j)}\alpha_{1}^{i+j}N^{q-2}-2\left(\alpha_{0}^{q-2-(k+l+\cdots)}\alpha_{1}^{k+l+\cdots}\chi_{ijk\ldots}\right)\,. 
\eea
We define the 2 $\times$ 2 matrices:
\bea
\label{gammaDefn}
T=\left(\begin{array}{cc}\tau & 0 \\ 0 & \tau\end{array}\right)\,,~~\Gamma=\left(\begin{array}{cc}\gamma_{00} & \gamma_{01} \\ \gamma_{10} & \gamma_{11}\end{array}\right)\,,~~ M=\gamma_{01}\left(\begin{array}{cc}v_2 & 1 \\ 1 & -v_1\end{array}\right)\,,
\eea
and the variables:
\bea
\label{vDefn}
\gamma_{01} v_1 := (\tau-\gamma_{11})\,,~~\gamma_{01} v_2 := (\gamma_{00}-\tau)\,,~~\gamma_{01}:=\gamma_{01}\,.
\eea
Using this notation we can write the Hessian in Eq. (\ref{hessGeneral}) as:
\bea
\label{hessGeneral2}
H = \left(\begin{array}{cccc} T & ~~\Gamma~~ & ~~\Gamma~~ ~~ & \cdots \\
~~\Gamma~~ & T & ~~\Gamma~~ ~~ & \cdots \\
~~\Gamma~~ & ~~\Gamma~~ & T ~~ & \cdots \\
~~\vdots~~ & ~~\vdots~~ & ~~\vdots~~  & \ddots 
\end{array}\right) = \left(\begin{array}{cccc} T & ~~(M+T)~~ & ~~(M+T)~~  & \cdots \\
~~(M+T)~~ & T & ~~(M+T)~~  & \cdots \\
~~(M+T)~~ & ~~(M+T)~~ & T  & \cdots \\
\vdots & \vdots & \vdots  & \ddots 
\end{array}\right)\,.
\eea

\section{Eigen-System of the Hessian}
\label{esystemSection}

Substituting (\ref{randyRule}) into the equations of motion (\ref{der1}) we obtain:
\bea
&& \alpha_i N^{q-1}-\chi_{ijk\cdots}\alpha_j\alpha_k\cdots=0\,,\nonumber\\
\label{sumEOM1}
i=0~~\to ~~ && \alpha_0 N^{q-1}-\chi_{00k\cdots}\alpha_0\alpha_k\cdots - \chi_{01k\cdots}\alpha_1\alpha_k\cdots=0\,,\nonumber \\
\label{chi}
~~\to~~ &&\chi_{00k\cdots}\alpha_k\cdots=\frac{1}{\alpha_0}(\alpha_0 N^{q-1}-\chi_{01k\cdots}\alpha_1\alpha_k\cdots)\,,\\
\label{sumEOM2}
i=1~~\to~~&&  \alpha_1 N^{q-1}-\chi_{10k\cdots}\alpha_0\alpha_k\cdots - \chi_{11k\cdots}\alpha_1\alpha_k\cdots=0\,,\nonumber\\
\label{chi2}
~~\to~~ && \chi_{11k\cdots}\alpha_k\cdots=\frac{1}{\alpha_1}(\alpha_1 N^{q-1}-\chi_{10k\cdots}\alpha_0\alpha_k\cdots)\,.
\eea

Similarly, substituting  (\ref{randyRule}) into (\ref{OFF}) we obtain:
\bea
\label{genGamma}
\gamma_{ij}&& =4\alpha_i \alpha_j N^{q-2}-2 \chi_{ijk\cdots}\alpha_k\alpha_l\cdots\,,\nonumber\\[2mm]
\label{gamma00}
\gamma_{00}&& =4\alpha_0^2 N^{q-2}-2\chi_{00k\cdots}\alpha_k\cdots\,, \\
\label{chi01}
\gamma_{01}&& =4\alpha_0\alpha_1 N^{q-2}-2\chi_{01k\cdots}\alpha_k\cdots ~~\to~~ \chi_{01k\cdots}\alpha_k\cdots=-\frac{1}{2}(\gamma_{01}-4\alpha_0\alpha_1 N^{q-2})\,,\\
\label{gamma11}
\gamma_{11}&& =4\alpha_1^2 N^{q-2}-2\chi_{11k\cdots}\alpha_k\cdots\,.
\eea

Substituting (\ref{chi}), (\ref{chi2}) and (\ref{chi01}) into (\ref{gamma00}) and (\ref{gamma11}) we obtain:
\bea
\label{gammaSUB}
\gamma_{00}&=&4\alpha_0^2 N^{q-2}-\frac{2}{\alpha_0}\big[\alpha_0 N^{q-1}+\frac{\alpha_1}{2}(\gamma_{01}-4\alpha_0\alpha_1 N^{q-2})\big]=\tau-\frac{\alpha_1}{\alpha_0}\gamma_{01}\,,\\
\label{gammaSUB2}
\gamma_{11}&=&4\alpha_1^2 N^{q-2}-\frac{2}{\alpha_1}\left[\alpha_1 N^{q-1}-\frac{\alpha_0}{2}(4\alpha_0\alpha_1 N^{q-2}-\gamma_{10})\right]= \tau-\frac{\alpha_0}{\alpha_1}\gamma_{10}\,.\nonumber
\eea
Using these expressions and (\ref{vDefn}) we obtain:
\bea
\label{vBasic}
v_1 = \frac{\alpha_0}{\alpha_1}\,,~~v_2 = -\frac{\alpha_1}{\alpha_0}\,,~~v_1\,v_2 = -1\,.
\eea

The result in (\ref{vBasic}) allows us to obtain analytic results for the eigenvectors and eigenvalues of the Hessian in Eq. (\ref{hessGeneral2}). The derivation is given in Appendix \ref{evvDerivation}, and the results are listed below. 
The eigenvectors have $2q$ components and we write them as lists of $q$ 2-component vectors:
\bea
\label{evectorsCase1}
&& V_1 = \big((v_1,1),(v_1,1),(v_1,1),\cdots\big)^T\,~~~~~~~~~~~~~~~~~{\rm eigenvalue} = q\tau \\[4mm]
&& \left.\begin{array}{l}
V_{2} = \big((-v_1,-1),(v_1,1),(0,0),\cdots\big)^T \\
V_{3} = \big((-v_1,-1),(0,0),(v_1,1),\cdots\big)^T \\
\vdots\\
V_{q} = \big((-v_1,-1),(0,0),\cdots(v_1,1)\big)\\
\end{array}\right\}~~~~~~~~~{\rm eigenvalue} = 0 \\[4mm]
&& \left.\begin{array}{l}
 V_{q+1} = \big((-v_2,-1),(v_2,1),(0,0),\cdots\big)^T\\
V_{q+2} = \big((-v_2,-1),(0,0),(v_2,1),\cdots\big)^T\\
\vdots\\
V_{2q-1} = \big((-v_2,-1),(0,0),\cdots(v_2,1)\big)
\end{array}\right\}~~~~~~{\rm eigenvalue} = \gamma_{01}(v_1-v_2)  \\[4mm]
\label{evectorsCase1f}
&& V_{2q} = \big((v_2,1),(v_2,1),(v_2,1),\cdots\big)^T\,~~~~~~~~~~~~~~~ 
{\rm eigenvalue}= q\tau-(q-1)\gamma_{01}(v_1-v_2)
\eea
We describe the content of this equation in words. 
\begin{enumerate}
\item The first eigenvector is called $V_1$. It has eigenvalue $q\tau$.

\item There are $(q-1)$ eigenvectors labelled $\{V_2,\cdots V_q\}$ which have $(-v_1,-1)$ in the first position, $(v_1,1)$ in any one of the remaining $(q-1)$ positions, and (0,0) in all remaining positions. All of these eigenvectors have eigenvalue 0. 

\item There are $(q-1)$ eigenvectors labelled $\{V_{q+1}\,\cdots V_{2q-1}\}$ with the same form as the eigenvectors $\{V_2,\cdots V_q\}$, except that $v_1$ is replaced by $v_2$. They have eigenvalues $\gamma_{01}(v_1-v_2)$. 

\item The last eigenvector is labelled $V_{2q}$ and has eigenvalue $q\tau-(q-1)\gamma_{01}(v_1-v_2)$.
\end{enumerate}

Using $v_1 v_2=-1$ (see Eq. (\ref{vBasic})), it is easy to see that all pairs of eigenvectors are orthogonal, except for pairs which have the same eigenvalue. We can construct a completely orthogonal set of eigenvectors using the slater determinant, in the usual way. \\

\section{Interpretation of Eigenvectors}
\label{interSection}

Now we consider the physical interpretation of the eigenvectors of the Hessian. We expect that these eigenvectors correspond to some kind of propagator-like normal modes.  We can represent the coefficients of the original product state as a six component vector:
\bea
\label{Vinitial}
V_{\rm initial} = (a_0,a_1,b_0,b_1,c_0,c_1,\cdots) = ((\alpha_0,\alpha_1),(\alpha_0,\alpha_1),(\alpha_0,\alpha_1)\cdots)^T\,.
\eea 
We consider translating this vector by an infinitesimal amount in the direction of each of the eigenvectors of the Hessian.  This produces a new vector:
\bea
\label{shifted}
V_i^\prime  = V_{\rm initial}+\epsilon V_i\,,~~i\in\{1,2,\dots q\}\,.
\eea

\ts

First we look at the eigenvector $V_1$. Using (\ref{vBasic}) we have:
\bea
\label{V1new}
&& V_1 = c\big((\alpha_0,\alpha_1),(\alpha_0,\alpha_1),(\alpha_0,\alpha_1)\cdots\big)^T\,,
\eea
and from (\ref{Vinitial}) and (\ref{shifted}) we obtain:
\bea
V_1^\prime 
= \big((\Lambda \alpha_0,\Lambda \alpha_1),(\Lambda \alpha_0,\Lambda \alpha_1),(\Lambda\alpha_0,\Lambda\alpha_1)\cdots\big)^T\,,~~~~\Lambda = (1+\epsilon c)\,.
\eea
We conclude that translating the product state by an infinitesimal vector in the direction of $V_1$ is equivalent to scaling the components of each qubit in the product state by $\Lambda$.

\ts

Next we look at the eigenvectors $\{V_2,\cdots V_q\}$. Using (\ref{vBasic}) the $V_2$ eigenvector can be written\footnote{To simplify the notation, we use the same letter for the constants in  (\ref{V1new}), (\ref{V2new}), (\ref{Vqnew}) and (\ref{V2qnew}).}
\bea
\label{V2new}
&& V_2 = c\big((-\alpha_0,-\alpha_1),(\alpha_0,\alpha_1),(0,0)\cdots\big)^T\,,
\eea
and from (\ref{Vinitial}) and (\ref{shifted}) we obtain:
\bea
V_{2}^\prime 
&&= ((\alpha_0(1-\epsilon c),\alpha_1(1-\epsilon c)),(\alpha_0(1+\epsilon c),\alpha_1(1+\epsilon c)),(\alpha_0,\alpha_1)\cdots)^T\,, \\
&&\approx ((\frac{1}{\lambda } \alpha_0,\frac{1}{\lambda } \alpha_1),(\lambda \alpha_0,\lambda \alpha_1),(\alpha_0,\alpha_1)\cdots)^T\,,~~~~\lambda = e^{-c\epsilon}\nonumber\,.
\eea
We conclude that translating the product state by an infinitesimal vector in the direction of $V_2$ is equivalent to scaling the components of the first qubit in the product state by $1/\lambda$, and the components of the second qubit  by $\lambda$. From Eq. (\ref{DsqDefn}), the distance measure is clearly invariant under this transformation. It is clear that the eigenvector $V_3$ can be treated in the same way, and corresponds to a scaling of the first and third qubits, and similarly for the eigenvectors $V_4$ to $V_q$. We conclude that the zero eigenvalue corresponds to the eigenvector that points in the direction of a symmetry of the distance function. 

\ts

Now we consider the eigenvectors $V_{q+1}$ to $V_{2q-1}$. Using (\ref{vBasic}) $V_{q+1}$ can be written: 
\bea
\label{Vqnew}
V_{q+1} = c((\alpha_1,-\alpha_0),(-\alpha_1,\alpha_0),(0,0)\cdots)\,,
\eea
and from (\ref{Vinitial}) and (\ref{shifted}) we obtain:
\bea
V_{q+1}^\prime 
&&= (R_{-c\epsilon}(\alpha_0,\alpha_1),R_{c\epsilon}(\alpha_0,\alpha_1),(\alpha_0,\alpha_1)\cdots)^T\,, \nonumber
\eea
where $R_\theta$ is the 2 $\times$ 2 matrix that generates a counter-clockwise rotation through an angle $\theta$: 
\bea
R_{\theta} = \left(\begin{array}{cc} \cos\theta & -\sin\theta \\ \sin\theta & \cos\theta\end{array}\right)\,.
\eea
We conclude that translating the product state by an infinitesimal vector in the direction of $V_{q+1}$ is equivalent to rotating the first qubit in the product state clockwise by an infinitesimal angle $\theta = \epsilon c$, and the second qubit in the product state counter-clockwise by the same angle. It is clear that the eigenvectors $V_{q+2}$ to $V_{2q-1}$ can be treated in the same way. 

\ts

Finally we consider the eigenvector $V_{2q}$. Using (\ref{vBasic}) it can be written: 
\bea
\label{V2qnew}
V_{2q} = c((-\alpha_1,\alpha_0),(-\alpha_1,\alpha_0),(-\alpha_1,\alpha_0)\cdots)\,,
\eea
and from (\ref{Vinitial}) and (\ref{shifted}) we obtain:
\bea
V_{2q}^\prime 
&&= (R_{c\epsilon}(\alpha_0,\alpha_1),R_{c\epsilon}(\alpha_0,\alpha_1),R_{c\epsilon}(\alpha_0,\alpha_1)\cdots)^T \,,\nonumber
\eea
which shows that translating the product state by an infinitesimal vector in the direction of $V_{2q}$ is equivalent to rotating each qubit in the product state counter-clockwise by an infinitesimal angle $\theta = \epsilon c$.

\ts

We conclude that the translation in Eq. (\ref{shifted}) has a simple interpretation, for each of the eigenvectors of the Hessian.

\section{Analytic Solution}
\label{AnaSection}

In this section we consider a general  class of permutation invariant target states that are consistent with the symmetries of the ansatz we are using, i.e. Eq. (\ref{randyRule}). We show that all eigenvalues of the Hessian (except the zero eigenvalues that correspond to a trivial scaling symmetry of the distance function) are positive, which means that the extremal solution is a local minimum, and therefore can be interpreted as a measure of the entanglement of the target state. In Appendix \ref{case3} we show that if we choose a target state that does not respect the symmetries of the original ansatz, it is not true that the extremum of the distance measure corresponds to a local minimum. This result shows that in order for the distance function to have a physical interpretation as a measure of entanglement, it is necessary to choose a target state that is consistent with the original ansatz, which is what we expect.

We construct a target state that is symmetric under the interchange of any 2 qubits by including all possible permutations of $p$ entries of ``1'' and $q-p$ entries of ``0''. We define the normalization factor:
\begin{equation}
\label{normFac}
{\cal A}^{-1}:= \sqrt{\left(\begin{array}{c} q \\ p\end{array}\right)}=\sqrt{\frac{q!}{p!(q-p)!}}\,\,.
\end{equation}
All non-zero values of $\chi_{ijk\dots}$ are equal to ${\cal A}$. The non-zero components correspond to the indices:
\bea
\label{genChi}
ijk\dots = {\cal P}(\underbrace{00\dots 0}_{q-p}\underbrace{11\dots 1}_{p})~~{\rm total~ number ~of ~terms ~is} \left(\begin{array}{c}q\\ p\end{array}\right)
\eea
For example, if $q=3$ and $p=1$, ${\cal A}=1/\sqrt{3}$ and the permutation invariant state is:
\bea
\label{psi symm}
\ket{\psi} = \frac{1}{\sqrt{3}}\left( \ket{0}\otimes \ket{0}\otimes \ket{1}+\ket{0}\otimes \ket{1}\otimes \ket{0}+\ket{1}\otimes \ket{0}\otimes \ket{0}\right)
\eea
The non-zero components are $\chi_{001}=\chi_{010}=\chi_{100}= 1/\sqrt{3}$.

Using (\ref{der1}) and (\ref{randyRule}) the extremal solution satisfies:
\bea
\label{soln4}
N^{q-1}={\cal A}\alpha_1^p \alpha_0^{q-p-2}\left(\begin{array}{c}q-1 \\ p\end{array}\right)\,,~~
N^{q-1}={\cal A}\alpha_1^{p-2}\alpha_0^{q-p}\left(\begin{array}{c}q-1 \\ p-1\end{array}\right)\,.
\eea
For the simple example in (\ref{psi symm} )these equations are:
\bea
N^2=\frac{2}{\sqrt{3}}\alpha_1 \, , ~~ N^2 = \frac{1}{\sqrt{3}}\frac{\alpha_0^2}{\alpha_1} 
\eea
These are readily solved to yield $N=\alpha_0^2+\alpha_1^2 =(4/9)^{1/3}$ and $\alpha_1^2= (3/4)N^4$, 

In general, rearranging the equations gives:
\bea
\label{genEOM}
\alpha _0^2 = N^X\left(\frac{q-p}{q}\right)^{Y+1}\left(\frac{q}{p}\right)^Z \left(\begin{array}{c}q-1 \\ p-1\end{array}\right)^W\,,\\
\alpha _1^2 = N^X\left(\frac{q-p}{q}\right)^Y\left(\frac{q}{p}\right)^{Z-1}\left(\begin{array}{c}q-1 \\ p-1\end{array}\right)^W\,,\nonumber
\eea
where we have defined the exponents:
\bea
X={2+\frac{2}{q-2}}\,,~~
Y={\frac{p-q}{q-2}}\,,~~
Z={\frac{p-1}{q-2}}\,,~~
W={\frac{1}{2-q}}\,.
\eea
Using these results it is straightforward to show:
\bea
\label{aDiv}
\frac{\alpha_0}{\alpha_1}=\sqrt{\frac{q-p}{p}}\,,~~~~~~\frac{N}{\alpha_0\alpha_1}=\sqrt{\frac{q-p}{p}} + \sqrt{\frac{p}{q-p}}\,.
\eea
{ These solutions can be used to obtain an analytic result for the distance measure \cite{passante}:
\bea
\label{distO}
D_c=1-N^q\,,~~N^q=
\left(\frac{p}{q}\right)^{p}\left(1-\frac{p}{q}\right)^{q-p}\left(\begin{array}{c}q\\p\end{array}\right)
\eea}
Note that the above agrees with the value obtained earlier for the simple case $q=3$, $p=1$.

Substituting (\ref{normFac}) and (\ref{genChi}) into (\ref{chi01}) we have:
\bea
\label{case4Help}
\gamma_{01} &=& 4\alpha_0\alpha_1 N^{q-2}-2\,\alpha_0^{q-p-1}\alpha_1^{p-1} {\cal A}\,\cdot\, \left(\begin{array}{c}q-2 \\ p-1\end{array}\right)\,,\\
&=& 4\alpha_0\alpha_1 N^{q-2}-2\,\frac{\sqrt{(q-p)p}}{q-1}N^{q-1}\,,\nonumber
\eea
where we have used (\ref{genEOM}) and (\ref{aDiv}) in the last line. 

Substituting  (\ref{vBasic}), (\ref{aDiv}) and (\ref{case4Help}) into (\ref{evectorsCase1})-(\ref{evectorsCase1f}) the eigenvalues are:
\bea
\label{evalueRes}
&& e_1 = q\tau ~~{\rm positive~ definite}\,,\\
&& e_2 = 0\,,\nonumber\\
&& e_3=\tau\left(1-\frac{1}{q-1}\right)~~{\rm positive~ definite ~for~} q>  2\,, \nonumber\\
&& e_4 = 2\tau~~{\rm positive~ definite}\,.\nonumber
\eea
The result is that all eigenvalues are positive definite for $q>2$ qubits, except for a set of zero eigenvalues that correspond to a trivial symmetry of the distance function. We conclude that the extremal solution corresponds to a  minimum.
The fact that the eigenvalues in (\ref{evalueRes}) are independent of $p$ is a consequence of restricting to a target state that does not mix values of $p$. A target state constructed as a linear combination of different values of $p$ would still satisfy the symmetry of the ansatz in (\ref{randyRule}), but the analytic solutions in this case are much more complicated. 

\section{Connection with Schmidt Decomposition}
\label{schmidtSection}

{ In this section we will show that the Schmidt decomposition of a general unnormalized product state yields a very convenient basis for the evaluation of the geometrical entanglement of an arbitrary target state.}
We start with the product state in Eq. (\ref{phi}). We construct a reduced density matrix by tracing over all qubits except the first:
\bea
M^A_{ii^\prime} = \phi_{ijk\dots}\phi_{i^\prime jk\dots} =\left[ N_B N_C\dots\right]\left(\begin{array}{cc}a_0^2 & a_0 a_1 \\ a_0 a_1 & a_1^2\end{array}\right)\,.
\eea
We find the eigenvectors $u_a^{(i)}$ and eigenvalues $\sigma_a^{(i)}$ for the reduced density matrix $M_A$:
 \bea && u_a^{(1)} = \left(\frac{a_0}{a_1},1\right)^T ~ {\rm with} ~ \sigma_a^{(1)} = N_A N_B N_C \dots\,, \\
&&   u_a^{(2)} = \left(-\frac{a_1}{a_0},1\right)^T ~ {\rm with} ~ \sigma_a^{(2)} = 0 \,.\nonumber
  \eea
Clearly these eigenvectors are orthogonal. We denote the normalized eigenvectors with hats and construct the unitary singular matrix: $A_{ji} = (\hat u_a^{(i)})_j$.

 We follow the same procedure for all other reduced density matrices. For example, tracing over all qubits but the second gives:
 \bea
&& M^B_{ii^\prime} = \phi_{ijk\dots}\phi_{ij^\prime k\dots} = \left[N_A N_C\dots\right]\left(\begin{array}{cc}b_0^2 & b_0 b_1 \\ b_0 b_1 & b_1^2\end{array}\right)\,,\\
 && u_b^{(1)} = \left(\frac{b_0}{b_1},1\right)^T ~ {\rm with} ~ \sigma_b^{(1)} = N_A N_B N_C \dots\,, \nonumber \\
&&   u_b^{(2)} = \left(-\frac{b_1}{b_0},1\right)^T ~ {\rm with} ~ \sigma_b^{(2)} = 0\,,\nonumber\\
\Rightarrow ~~~~~~~&& B_{ji} = (\hat u_b^{(i)})_j\,.\nonumber
\eea
We construct the singular value decomposition (SVD) of the { matrix representation of the product state implied by Eq. (\ref{phi})}:
\bea
\label{tildePhi}
\phi_{ijk\dots}=  \left[A_{ix}B_{jy}C_{kz}\dots\right]\tilde\Sigma_{xyz\dots}\,,~~A = \frac{1}{\sqrt{N_A}}\left(\begin{array}{cc}a_0 & -a_1 \\ a_1 & a_0 \end{array}\right)\,,~~B = \frac{1}{\sqrt{N_B}}\left(\begin{array}{cc}b_0 & -b_1 \\ b_1 & b_0 \end{array}\right)\,~~\dots
\eea
It is easy to show that:
\bea
\label{tildeSigma}
\tilde \Sigma_{ijk\dots}=\left[A^{-1}_{ix} B^{-1}_{jy}C^{-1}_{kz} \dots\right]\phi_{xyz\dots} = \sigma\,\delta_{1i}\delta_{1j}\delta_{1k}\dots\,,~~~
  \sigma:=\tilde\Sigma_{111\dots}=\sqrt{N_A N_B N_C} \cdots\,.
 \eea
 { For example, for a 2 qubit state, using (\ref{phi}) and (\ref{tildePhi}) we have:
 \bea
\phi_{ij}=\left(\begin{array}{cc} a_0 b_0 & a_0 b_1 \\ a_1 b_0 & a_1 b_1 \end{array}\right)\,,~~ 
 A^{-1}=\frac{1}{\sqrt{N_A}}\left(\begin{array}{cc} a_0 & a_1\\-a_1& a_0 \end{array}\right)\, , ~~ B^{-1}=\frac{1}{\sqrt{N_B}}\left(\begin{array}{cc} b_0 & b_1\\-b_1& b_0 \end{array}\right)\,,.
 \eea
 Substituting into (\ref{tildeSigma}) we  obtain:
 \bea
\tilde \Sigma=\sqrt{N_a N_b}\left(\begin{array}{cc} 1 & 0\\0 & 0 \end{array}\right)\,.
 \eea}
 
 We can write the target state in the same basis:
 \bea
 \label{psiSigma}
&& \chi_{ijk\dots}= \left[A_{ix}B_{jy}C_{kz}\dots\right] \Sigma_{xyz\dots}\,, \\[2mm]
&&  \Sigma_{xyz\dots}=\left[A^{-1}_{xi} B^{-1}_{yj} C^{-1}_{zk} \dots\right]\chi_{ijk\dots}\,.\nonumber
\eea
 The matrix $\Sigma$ is messy, but we don't need to use the form of $\Sigma$ to establish the connection between the geometric definition of entanglement in section \ref{sectionIntro} and the Schmidt decomposition.

 From Eqs. (\ref{DsqDefn}) and (\ref{psiSigma}) we have:
 \bea
 D^2 = \langle \psi-\phi| \psi-\phi\rangle = (\Sigma-\tilde\Sigma)\cdot (\Sigma-\tilde\Sigma)\,,
 \eea
 where we have defined $M\cdot N := M_{ijk\dots}N_{ijk\dots}$ for arbitrary tensors $M$ and $N$.
 Using (\ref{tildeSigma}) we have:
 \bea
 D^2 = 1+\sigma^2-2\sigma\Sigma_{111\dots}\,,
 \label{Dsquared}
 \eea
 where we have used $\langle\psi|\psi\rangle = \Sigma\cdot\Sigma =1$ since the target state is assumed to be normalized. 
 
 We can find the value of $\sigma$ that minimizes this distance by solving:
 \bea
 \label{deriv1}
   \frac{1}{2}\frac{d}{d\sigma} D^2\bigg|_{\sigma=\sigma_c} = 0\,.
   \eea
First we show that $\Sigma_{111\dots}$ is independent of $\sigma$. It is simple to see this in polar co-ordinates:
\bea
a_0=a\cos\theta_a\,,~a_1=a\sin\theta_a\,;~b_0=b\cos\theta_b\,,~b_1=b\sin\theta_b\,;~c_0=c\cos\theta_c\,,~c_1=c\sin\theta_a\,\dots
\eea
Using the notation:
$
i_1=a\,,~~i_2=b\,,~~i_3=c\,,\dots
$
the distance function has the form (see Eq. (\ref{DsqDefn})):
\bea
\label{DsqPolar}
D^2 = 1-2 \sqrt{q} \sigma \sum_{x=1}^q \sin (\theta_{i_x} ) \prod_{j=1|j\ne x}^{q-1} \cos (\theta_{i_j} )+\sigma^2\,.
\eea
The middle term on the rhs of this expression is a sum of terms of the form $(\sin\theta_a\cos\theta_b\cos\theta_c\cos\theta_d\dots)$ + $(\cos\theta_a\sin\theta_b\cos\theta_c\cos\theta_d\dots) + \dots$. In each term there is 1 sine factor and  $(q-1)$ cosine factors, and there are $q$ terms which correspond to $q$ different choices for the location of the lone sine factor. 
The point is that (\ref{DsqPolar}) depends on the $q+1$ variables $ \theta_{i_j}$ $(j\in\{1,2,\dots q\})$ and $\sigma$, in contrast to (\ref{DsqDefn}), which depends on the 2$q$ variables $\{a_0,a_1,b_0,b_1\dots\}$. 
The Hessian is a $(q+1)\times(q+1)$ dimensional matrix that does not have zero eigenvalues, since the symmetry which produces the zero eigenvalues (see section \ref{interSection}) has been removed by the switch to polar co-ordinates. We show this explicitly in Appendix \ref{polarSection}. 

In polar co-ordinates the matrices in (\ref{tildePhi}) have the form:
\bea
\label{Apolar}
A = \left(\begin{array}{cc}\cos\theta_a & -\sin\theta_a \\ \sin\theta_a & \cos\theta_a \end{array}\right)\,,~~B = \left(\begin{array}{cc}\cos\theta_b & -\sin\theta_b \\ \sin\theta_b & \cos\theta_b \end{array}\right)\,,\dots
\eea
From (\ref{psiSigma}) and (\ref{Apolar}) it is clear that $\Sigma_{111\dots}$ is independent of $\sigma$.
Using this result (\ref{Dsquared}) and (\ref{deriv1}) becomes:
 \bea
 \label{minEqn}
 \frac{1}{2}\frac{d}{d\sigma} D^2 = \sigma-\Sigma_{111\dots}\bigg|_{\sigma=\sigma_c} = 0~~\to~~\sigma_c=\Sigma_{111\dots}\,,
 \eea
 which gives:
 \bea
  \label{crit}
 D^2_c  =  1-\sigma_c^2\,.
 \eea
 To compare with the geometric distance measure we look at the angle between $\phi$ and $\psi$ which is defined in the second line of Eq. (\ref{cangle}):
 \bea
 \cos\theta= \frac{~~\langle\phi|\psi\rangle}{\sqrt{\langle\phi|\phi\rangle}\sqrt{\langle\psi|\psi\rangle}}=\frac{~~~\tilde\Sigma \cdot \Sigma}{\sqrt{\tilde\Sigma\tilde\Sigma}} = \frac{\sigma \Sigma_{111\dots}}{\sqrt{\sigma^2}}~~~~\to~~~~\cos\theta_c = \sigma_c\,,
 \eea
 and thus we have 
 \bea
  D^2_c = 1-\cos^2\theta_c= 1-\Sigma^2_{111...}\,,
  \eea
  which agrees with the result in the first line of Eq. (\ref{cangle}).
  
Recall that in order to complete the calculation of the geometrical entanglement, it is necessary to extremize $\Sigma_{111...}$ with respect to the remaining parameters describing the unnormalized product states. Using polar coordinates these are the angles $\theta_a,\theta_b...$ that appear in (\ref{Apolar}). While this will yield a complicated set of non-linear equations in general, in the case of the permutation symmetric target and product states of the previous sections things simplify considerably. In particular, for the product states to be permutation invariant, all the angles must be the same:
\bea
\theta:=\theta_a=\theta_b=... \, .
\label{randyRulePolar}
\eea 
Given Eq.(\ref{psiSigma}) and the form (\ref{genChi}) of $\chi_{ijk...}$ in the case of a permutation invariant target state, we have that:
\bea 
\label{Sigma minimization}
\Sigma_{111...}
  &=& \frac{1}{{\cal A}}(\cos(\theta))^{q-p}(\sin(\theta))^p 
\eea
The extremal condition $\partial \Sigma_{111...}/ \partial \theta =0$ then yields the solution:
\bea
\label{theta solution}
\tan^2({\theta_c})= \frac{p}{q-p} 
\eea 
Substituting this back into $\Sigma_{111...}$ gives the minimum distance of $1-\sigma^2_c$ with:
\bea
\label{sigma solution}
\sigma_c^2=N^q=\Sigma_{111...}^2 =  \left(\begin{array}{c}q\\ p\end{array}\right) \left(\frac{p}{q}\right)^p \left(1-\frac{p}{q}\right)^{(q-p)}
\eea
in agreement with Eq.(\ref{distO}).

For clarity we illustrate the above using the example in Eqs.(\ref{psi symm}) of the previous section, namely $q=3$ and $p=1$. In this case we have:
\bea
\label{Bipolar}
A^{-1} = B^{-1}= C^{-1} =\left(\begin{array}{cc}\cos\theta & \sin\theta \\ -\sin\theta & \cos\theta \end{array}\right) \,\, .
\eea
Thus,
\bea
\Sigma_{111}&=& A^{-1}_{1i} A^{-1}_{1j} A^{-1}_{1k} \chi_{ijk}\nonumber\\
  &=& \cos\theta \, \cos\theta \, \sin\theta \,\chi_{001} + \cos\theta\,\sin\theta\,\cos\theta \,\chi_{010}+\sin\theta\,\cos\theta\,\cos \theta\, \chi_{100}\nonumber\\
  &=& 3\times\frac{1}{\sqrt{3}}\cos^2\theta\,\sin\theta
\label{sigma example}
\eea
as in (\ref{Sigma minimization}). Extremizing this expression with respect to variations of $\theta$ yields $\sigma_c^2=4/9$, as expected from the general solution (\ref{sigma solution}).
  \section{Conclusions}
  \label{concSection}
  \label{concSECT}
We have studied a generalisation of the usual
geometric measure of entanglement of pure states using the
distance to the nearest unnormalised product state. 
When the target state has a large degree of symmetry, one can find general examples for which one can solve the system of non--linear equations analytically.  For these solutions, we have proven that all eigenvalues of the Hessian are positive, which means that the extrema are local minima. This provides a local (as opposed to global) version for unnormalized product states of the proof by Hubener {\it et al} \cite{hubener} for normalized product states that permutation symmetric product states minimize the distance to permutation symmetric target states.  
In addition, we have shown that the conditions that determine the extremal solutions for general target states can be obtained directly by parametrizing the product states via their Schmidt decomposition. 

Our results verify that the distance measure we have defined in \cite{passante} is a meaningful measure of entanglement. In order to show that it is potentially useful for physical systems, one needs to analyze it in more general settings such as multi-partite target states with less symmetry and mixed, as opposed to pure, states. This work is currently in progress. \\[10pt]

\noindent
 {\bf Acknowledgements}
This paper is dedicated to the memory of our collaborator Randy Kobes, who passed away in September, 2010.
We gratefully acknowledge the support of Natural Sciences and Engineering Research Council of Canada. G. Kunstatter thanks Robert James for useful discussions.

\appendix

\section{Derivation of the eigenvectors and eigenvalues of the Hessian}
\label{evvDerivation}

It is straightforward to verify the eigenvectors and eigenvalues in Eqs. (\ref{evectorsCase1}) - (\ref{evectorsCase1f}). The eigenvalue equation can be written:
\bea
{\rm row}_n \cdot V_m = E_m\,V_m[n]\,,
\eea
where row$_n$ is the $n$-th row of the Hessian, $E_m$ is the $m$-th eigenvalue, and $V_m[n]$ is the $n$-th component of the $m$-th eigenvector. Because of the symmetry of the Hessian and eigenvectors, we only need to look at the first 2 rows, and the eigenvectors $V_1$, $V_2$, $V_{q+1}$ and $V_{2q}$. We write out the first two rows of the Hessian in Eq. (\ref{hessGeneral2}):
\bea
&& {\rm row}_1 = \big((\tau,0),(\gamma_{01} v_2+\tau,\gamma_{01}),(\gamma_{01} v_2+\tau,\gamma_{01}),(\gamma_{01} v_2+\tau,\gamma_{01})\cdots\big)\,, \\
&& {\rm row}_2 = \big((0,\tau),(\gamma_{01},-\gamma_{01} v_1+\tau),(\gamma_{01},-\gamma_{01} v_1+\tau),(\gamma_{01},-\gamma_{01} v_1+\tau)\cdots\big)\,.\nonumber
\eea
Contracting these rows with the eigenvectors $V_1$, $V_2$, $V_{q+1}$ and $V_{2q}$ we obtain:
\bea
{\rm row}_1 V_1&& = (v_1,0)\cdot (\tau,0)^T + (q-1)\big[(v_1,1)\cdot(\gamma_{01} v_2+\tau,\gamma_{01})^T\big] \\
&&= q\tau v_1+(q-1)\gamma_{01}(1+v_1v_2) = q\tau(v_1)=E_1\,V_1[1]\,,\nonumber\\[2mm]
{\rm row}_2 V_1 &&= (v_1,0)\cdot(0,\tau)^T + (q-1)\big[(v_1,1)\cdot(-\gamma_{01} v_1+\tau,\gamma_{01})^T\big] = q \tau = E_1\, V_1[2]\,, \nonumber\\[2mm]
{\rm row}_1 V_2 && = (\tau,0)\cdot(-v_1,-1)^T+(\gamma_{01} v_2+\tau,\gamma_{01})\cdot(v_1,1)^T = \gamma_{01}(1+v_1 v_2) = 0=E_2\,V_2[1]\,, \nonumber\\[2mm]
{\rm row}_2 V_2 &&  = (0,\tau)\cdot(-v_1,-1)^T+(-\gamma_{01} v_1+\tau,\gamma_{01})\cdot(v_1,1)^T = 0=E_2\,V_2[2]\,, \nonumber\\[2mm]
{\rm row}_1 V_{q+1} && = (\tau,0)\cdot(-v_2,-1)^T + (\gamma_{01} v_2+\tau,\gamma_{01})\cdot(v_2,1)^T = \gamma_{01}(1+v_2^2) \nonumber\\
&&= \gamma_{01}(v_1-v_2)(-v_2)= E_{q+1}\,V_{q+1}[1]\,, \nonumber\\[2mm]
{\rm row}_2 V_{q+1} && = (0,\tau)\cdot(-v_2,-1)^T + (-\gamma_{01} v_1+\tau,\gamma_{01})\cdot(v_2,1)^T = \gamma_{01}(v_1-v_2)(-1)= E_{q+1}\, V_{q+1}[2]\,, \nonumber\\ [2mm]
{\rm row}_1 V_{2q} && = (v_2,0)\cdot (\tau,0)^T + (q-1)\big[(v_2,1)\cdot(\gamma_{01} v_2+\tau,\gamma_{01})^T\big] \nonumber\\
&&= q\tau v_2+(q-1)\gamma_{01}(1+v_2^2) = \big(q\tau-(q-1)\gamma_{01}(v_1-v_2)\big)(v_2)=E_{2q}\,V_{2q}[1]\,,\nonumber\\[2mm]
{\rm row}_2 V_{2q} && = (v_2,0)\cdot(0,\tau)^T+(q-1)\big[(v_2,1)\cdot(-\gamma_{01} v_1+\tau,\gamma_{01})^T\big] \nonumber\\
&&= (q\tau -(q-1)\gamma_{01}(v_1-v_2)) = E_{2q}\,V_{2q}[2]\,,\nonumber
\eea

\section{Symmetry Violating Example}
\label{case3}

Consider the target state: 
\bea
\!\!\!\!\!\ket{\psi} =  \left(
\, \ket{1\,1\,0\,0\,\ldots 0} +\, \ket{0\,1\,1\,0\,\ldots 0}+\, \ket{0\,0\,1\,1\,\ldots 0} +\ldots + \ket{0\,0\,\ldots\,1\,1}+ \ket{1\,0\,\ldots\,0\,1}\right)\,\frac{1}{\sqrt{q}}\,.
\eea 
This target state does not correspond to the state in section \ref{AnaSection} with $p=2$, because in this case $\chi_{ij\dots}$ is non--zero only when there are two adjacent 1's (using periodic `boundary conditions'):
\begin{equation}
\label{norm3}
\chi_{11000\cdots 0} = \chi_{01100\cdots 0}  = \chi_{00110\cdots 0} =\chi_{00011\cdots 0}= \cdots=\chi_{00000\cdots 11} =\chi_{10000\cdots 01}=
\frac{1}{\sqrt{q}}\,.
\end{equation}
The target state is not symmetric under the interchange of any two qubits, and therefore it does not respect the symmetry of the ansatz in Eq. (\ref{randyRule}).
Using  (\ref{der1}) and (\ref{randyRule}) the extremal solution satisfies:
\bea
\label{soln3}
N^{q-1}=\frac{(q-2)}{\sqrt{q}} \alpha _1^2 \alpha _0^{q-4}\,,~~
N^{q-1}=\frac{2}{\sqrt{q}} \alpha _0^{q-2}\,.
\eea
Rearranging gives:
\bea
\label{rearange}
\alpha _0^2 =  \frac{N (q-2)}{q}\,,~~\alpha _1^2 = \frac{2 N}{q}\,.
\eea
Using (\ref{off}) we have: 
\bea
\label{case3Help}
\gamma_{01}&=&4\alpha_0\alpha_1 N^{q-2}-\frac{2}{\sqrt{q}}\alpha_0^{q-3}\alpha_1\,,\\\nonumber &=& 4\alpha_0\alpha_1 N^{q-2}-\frac{\alpha_1}{\alpha_0}N^{q-1} = 4\alpha_0\alpha_1 N^{q-2}-\sqrt{\frac{2}{q-2}}N^{q-1}\,,
\eea
where we used (\ref{norm3}) in the first line, and (\ref{soln3}) and (\ref{rearange}) in the second line. 
Substituting  (\ref{vBasic}), (\ref{rearange}) and (\ref{case3Help}) into  (\ref{evectorsCase1})-(\ref{evectorsCase1f}) the eigenvalues are:
\bea
&& e_1 = q\tau ~~{\rm positive~ definite}\,,\\
&& e_2 = 0\,,\nonumber\\
&& e_3=\tau\left(2-\frac{q}{2(q-2)}\right)~~{\rm positive~ definite ~for~} q\ge 3\,, \nonumber\\
&& e_4 = -\tau\left(\frac{q^2-7q+8}{2(q-2)}\right)~~{\rm negative~ definite~for~}q\ge 6\,.\nonumber
\eea
The existence of a negative eigenvalue means that the extremal solution is not a local minimum. This result is not unexpected, since the target state is not consistent with the symmetry required by the ansatz in Eq.  (\ref{randyRule}).

\section{The Hessian in Polar co-ordinates}
\label{polarSection}

Using the solution (\ref{sigma solution}), defining $\mathbb{N}\equiv N^q$ and 
 setting $p=1$ for simplicity we obtain:
\bea
\label{eomPolar}
\mathbb{N}=\left(\frac{q-1}{q}\right)^{q-1}\,,~~~\theta =\cos ^{-1}\left(\sqrt{1-\frac{1}{q}}\right)\,,
\eea
which reproduces $D^2_{min}=1-\mathbb{N}$, in agreement with (\ref{tildeSigma}) and (\ref{crit}).

The Hessian has the form:
\bea
&& H=
\left[\begin{array}{cccccc} B & Z & Z & Z & Z &\dots \\ Z & M & X & X & X &\dots  \\ Z & X & M & X & X &\dots \\ Z & X & X & M & X &\dots  \\ \vdots &\vdots &\vdots &\vdots &\vdots &\ddots\end{array}\right] 
\eea
\bea
\label{polarEV}
B =\frac{\partial^2}{\partial^2 \mathbb{N}}D^2&& ~~ \to~~  \frac{\sqrt{q}}{2 \mathbb{N}^{3/2}} \sin (\theta ) \cos ^{q-1}(\theta )~~  \to~~ 2\mathbb{N}  \\
 Z=\frac{\partial^2}{\partial \mathbb{N}\partial \theta_i}D^2 &&~~\to~~ -\frac{1}{\sqrt{q} \sqrt{\mathbb{N}}}\cos ^q(\theta )+(1-q) \sin ^2(\theta ) \cos ^{q-2}(\theta )~~\to~~0 \nonumber \\
 M =\frac{\partial^2}{\partial \theta_i\partial \theta_i}D^2&& ~~\to~~ 2 \sqrt{q} \sqrt{\mathbb{N}} \sin (\theta ) \cos ^{q-1}(\theta )~~\to~~\frac{1}{2\mathbb{N} } \nonumber\\
  X =\frac{\partial^2}{\partial \theta_i\partial \theta_j} D^2 &&~~\to~~ -2\frac{\sqrt{\mathbb{N}}}{\sqrt{q}} \left((q-2) \cos ^2(\theta ) \sin ^{q-2}(\theta )-2 \sin (\theta ) \cos ^{q-1}(\theta )\right) \nonumber\\
  &&~~\to~~ \frac{2}{q} \left(2-(q-2) (q-1)^{\frac{3}{2}-\frac{q}{2}}\right) \mathbb{N} \,.\nonumber
\eea
In each line of (\ref{polarEV}), the first arrow indicates the results obtained using the ansatz (\ref{randyRulePolar}), and the second arrow indicates that we have used the equation of motion (\ref{eomPolar}). It is straightforward to calculate the eigenvalues:
\bea
\label{evPolar}
e_1 = B\,,~~e_2=e_3=\dots = e_{q} = M-X\,,~~e_{q+1} = M+4X\,.
\eea
As expected, the zero eigenvalues have disappeared, and we have 3 distinct eigenvalues, as before.

It is straightforward to see how the eigenvalues in (\ref{evPolar}) are related to those in (\ref{evectorsCase1})-(\ref{evectorsCase1f}). We look at one example. Using the chain rule we have:
\bea
\frac{\partial^2}{\partial \mathbb{N}^2}D^2 = &&
\frac{\partial\alpha_0}{\partial\mathbb{N}} q \left(\frac{\partial\alpha_0}{\partial\mathbb{N}} (q-1) \gamma _{00}+\frac{\partial\alpha_1}{\partial\mathbb{N}} (q-1) \gamma _{01}  +\frac{\partial\alpha_0}{\partial\mathbb{N}} \tau \right) \\
&& +\frac{\partial\alpha_1}{\partial\mathbb{N}} q \left(\frac{\partial\alpha_0}{\partial\mathbb{N}}
   (q-1) \gamma _{01}+\frac{\partial\alpha_1}{\partial\mathbb{N}} (q-1) \gamma _{11}+\frac{\partial\alpha_1}{\partial\mathbb{N}} \tau \right)\,.\nonumber
   \eea
   The left-hand side of this equation is the eigenvalue $e_1$ in (\ref{evPolar}). The right-handside is a function of the eigenvalues in (\ref{evectorsCase1})-(\ref{evectorsCase1f}). 
   Using (\ref{gammaSUB}) and (\ref{gammaSUB2}) we obtain:
   \bea
  \frac{1}{q} \frac{\partial^2}{\partial \mathbb{N}^2}D^2 = 
\frac{\partial\alpha_0}{\partial\mathbb{N}}^2 \left(q \left(v_2 \gamma _{01}+\tau \right)-v_2 \gamma _{01}\right)+2 \frac{\partial\alpha_0}{\partial\mathbb{N}} \frac{\partial\alpha_1}{\partial\mathbb{N}} (q-1)
   \gamma _{01}+\frac{\partial\alpha_1}{\partial\mathbb{N}}^2 \left(q \left(\tau -v_1 \gamma _{01}\right)+v_1 \gamma _{01}\right)\,.\nonumber
   \eea
   The derivatives can be calculated directly:
   \bea
   \frac{\partial\alpha_0}{\partial\mathbb{N}} = \frac{\sqrt{q-1} }{2 q^{3/2}} \mathbb{N}^{\frac{1}{2 q}-1}\,,~~\frac{\partial\alpha_1}{\partial\mathbb{N}} = \frac{1}{2 q^{3/2}}\mathbb{N}^{\frac{1}{2 q}-1}\,.
   \eea
 It is straightforward to show that these results satisfy:
 \bea
 \frac{\partial\alpha_0}{\partial\mathbb{N}} = v_1 \frac{\partial\alpha_1}{\partial\mathbb{N}}\,,~~\frac{\partial\alpha_1}{\partial\mathbb{N}}^{-2} = 2q^2 v_1(v_1-v_2)\tau\mathbb{N}\,.
 \eea
 Substituting we reproduce the first equation in (\ref{evPolar}).

\end{document}